\documentclass{article}
\usepackage{spconf}
\usepackage{amsmath,amsfonts,amssymb,amsthm}
\usepackage{graphicx}
\usepackage{float}
\usepackage{subcaption}
\usepackage{algorithm,algpseudocode} 
\usepackage{lipsum}

\newcommand\blfootnote[1]{%
  \begingroup
  \renewcommand\thefootnote{}\footnote{#1}%
  \addtocounter{footnote}{-1}%
  \endgroup
}

\title{Patient-adaptive and Learned MRI Data Undersampling Using \\ Neighborhood Clustering}

\name{Siddhant Gautam$^{\star}$ \qquad Angqi Li$^{\star}$ \qquad Saiprasad~Ravishankar$^{\star \dagger}$\thanks{S. Gautam (\textit{gautamsi@msu.edu}) and S. Ravishankar (\textit{ravisha3@msu.edu}) are the corresponding authors. All authors were supported in part by the NIH grant R21 EB030762.}}
  
\address{$^{\star}$ Dept. of Computational Mathematics Science and Engineering,\\Michigan State University, East Lansing, MI, USA \\
  $^{\dagger}$ Dept. of Biomedical Engineering, Michigan State University, East Lansing, MI, USA}

\begin{document}
\ninept

\maketitle

\begin{abstract}
 There has been much recent interest in adapting undersampled trajectories in MRI based on training data. In this work, we propose a novel patient-adaptive MRI sampling algorithm based on grouping scans within a training set. Scan-adaptive sampling patterns are optimized together with an image reconstruction network for the training scans. The training optimization alternates between determining the best sampling pattern for each scan (based on a greedy search or iterative coordinate descent (ICD)) and training a reconstructor across the dataset. The eventual scan-adaptive sampling patterns on the training set are used as labels to predict sampling design using nearest neighbor search at test time. The proposed algorithm is applied to the fastMRI knee multicoil dataset and demonstrates improved performance over several baselines.
\end{abstract}

\blfootnote{$\copyright$ 2024 IEEE.  Personal use of this material is permitted.  Permission from IEEE must be obtained for all other uses, in any current or future media, including reprinting/republishing this material for advertising or promotional purposes, creating new collective works, for resale or redistribution to servers or lists, or reuse of any copyrighted component of this work in other works.}


%
\vspace{-0.1in}
\section{Introduction}
Reconstructing MR images from undersampled measurements has been of considerable interest to the medical imaging community. Reducing scan time for MRI acquisition offers many benefits, including reducing patient discomfort, increasing scaling throughput, and alleviating motion and other artifacts.

Pulse sequence and k-space trajectory design~\cite{POSER2018, Bitar_06_MRI_review}, and partial or half Fourier methods~\cite{Liang_Book} have been used to improve sampling efficiency. Parallel MRI (p-MRI)~\cite{pMRI-Survey} was proposed to give hardware-based acceleration but it suffers from increased noise and imperfect artifact correction at higher undersampling rates.




Compressive sensing-based techniques have been proposed in the past to recover MR images from sparse or subsampled measurements \cite{lustig2007sparse, emmanuel2004robust}. In contrast to classical compressive sensing (CS), which presupposes familiarity with the sparsity bases of signals or incoherence between the basis and the measurement operator, approaches using learned image models for reconstruction have demonstrated greater efficacy. These methods include techniques like (patch-based) synthesis dictionary learning~\cite{ravishankar2010mr, lingala2013blind}. Additionally, recent progress in transform learning~\cite{ravishankar2012learning,ravishankar2019image} presents an alternative framework for achieving 
efficient sparse modeling in MRI.

With the advent of deep learning,
the U-Net~\cite{ronneberger2015u} and its variants~\cite{zhou2018unet++, lee2018deep} have been quite successful in biomedical image segmentation and reconstruction. Unrolled methods such as model-based deep learning (MoDL)~\cite{aggarwal2018modl} 
use the MRI forward model within a data consistency step and a CNN reconstructor as a denoiser to regularize the reconstruction.

Recently, deep learning-based approaches have invited considerable interest for sampling prediction in accelerated MRI, including methods such as LOUPE~\cite{bahadir2020deep, xie2022puert}. Some works have jointly learned sequential sampling along with a reconstruction strategy \cite{jin2017escape}. However, a limitation of approaches such as LOUPE, J-MoDL \cite{J-MoDL}, and the method of~\cite{sherry2020} is that the learned sampling pattern is not adapted to every test case in a patient or scan adaptive manner, but rather fixed after it is learned on the training set. In this sense, LOUPE learns a sampling pattern that is adapted to the entire dataset (population-adaptive) rather than to individual scans. Scan-adaptive sampling prediction could lead to potentially improved sampling efficiency and reconstruction performance compared to population-adaptive methods.




While most previous approaches focused on population adaptive sampling for MRI \cite{bahadir2020deep, yin2021end}, in this work, we focus on learning 
patient and scan adaptive MRI sampling patterns. The optimal sampling patterns are jointly estimated along with a deep reconstructor network. Our previous work explored scan-adaptive mask-prediction networks for single-coil MRI data~\cite{huang2022single}. However, it suffers from drawbacks due to a lack of training labels and this problem is addressed in this work.



We propose a training optimization that finds scan-adaptive sampling patterns and learned reconstruction models for multi-coil MRI. Our algorithm can alternatingly estimate a reconstructor and a collection of sampling patterns from training data (Section~\ref{sec:alternate_framework}). The novel contributions include generating patient and scan adaptive sampling patterns whereas earlier approaches have mostly designed population-adaptive masks~\cite{bahadir2020deep, gozcu2018learning}. We use a greedy algorithm~\cite{gozcu2018learning} to generate initial high-quality sampling patterns which are then used as starting points for mask optimization and are further refined using an iterative coordinate descent scheme to yield improved sampling masks on training data. The details of mask optimization are provided in Section~\ref{sec:greedy_icd_algorithm}. At testing time, a nearest neighbor search is used to estimate the high-frequency information to sample (from the pre-learned patient-adaptive masks) based on acquired low-frequency information in the test scan. Section~\ref{sec:nn_search} discusses in detail how the nearest neighbor search is used to obtain scan-adaptive masks at testing time and Section~\ref{sec3} demonstrates the performance of the learned masks.


\section{Methodology}
\subsection{Multi-coil MRI Reconstruction}
The multi-coil MR image reconstruction problem from subsampled k-space measurements often takes the form of a regularized optimization problem:
\begin{equation}
    \underset{\mathbf{x}}{\arg\min} \sum_{c=1}^{N_c} \| \mathbf{A}_c \mathbf{x}-\mathbf{y}_c\|_2^2 + \lambda R(\mathbf{x}),
\end{equation}
where $\mathbf{A}_c= \mathbf{M} \mathbf{F} \mathbf{S}_c$ is the forward MRI operator and $R(\mathbf{x})$ is a regularizer. 
In particular,
$\mathbf{M}$ is a sampling operator that subsamples k-space, $\mathbf{F}$ is the Fourier transform operator, $\mathbf{S}$ is the coil sensitivity encoding operator, $N_c$ is the number of coils, and $\mathbf{x}$ is the reconstructed image. 
The regularizer typically captures assumed properties of the image such as sparsity in a transform domain, low-rankness, etc.





\subsection{Framework for Jointly Learning Reconstructor and Sampler} \label{sec:alternate_framework}
Using a training set consisting of fully-sampled k-space scans, we learn scan-adaptive sampling masks and a reconstructor from the training data.
We learn scan-adaptive sampling masks $\mathbf{M}_i$ and a reconstructor $f_\theta$
by alternating minimization that optimizes the following problem:
\begin{equation}\label{eq:opt}
    \underset{\theta, \left\{ \mathbf{M}_i \in \mathcal{C}  \right\} }{\min} \sum_{i=1}^N \| f_\theta (\mathbf{A}_i^H \mathbf{M}_i (\mathbf{y}_i) ) - \mathbf{x}_i \|_2^2,
\end{equation}
where $\mathbf{M}_i$ is the $i$th training mask that inserts zeros at non-sampled locations, $\mathbf{y}_i$ and $\mathbf{x}_i$ are the $i$th fully-sampled training k-space and the corresponding ground truth image, respectively, $\mathbf{A}_i^H$ is the adjoint of the fully-sampled multicoil MRI measurement operator for the $i$th training scan, and $f_\theta$ is the reconstruction network. $\mathcal{C}$ is the set of all sampling masks that zero all but a subset of k-space, with a specified sampling budget.

\begin{figure}[ht]
    \centering
    \includegraphics[width=0.8\linewidth]{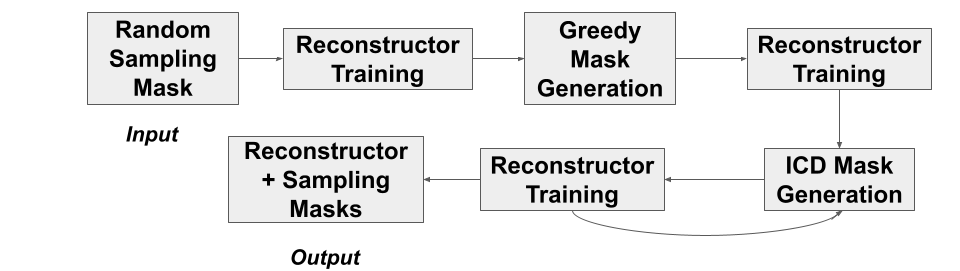}
    \caption{Alternating framework for mask and reconstructor update during joint training.}
     \label{fig:maintraining}
\end{figure} 

Figure~\ref{fig:maintraining} shows our algorithm for jointly learning scan-adaptive sampling masks and a reconstructor on training data. It updates the reconstructor and sampling masks alternatively through different iterations.
In the first pass of training, the aliased images (referring to $\mathbf{A}^H \mathbf{M}_i (\mathbf{y}_i)$) are generated from a conventional variable density random sampling (VDRS) mask~\cite{lustig2007sparse}. Given the sampling patterns $\mathbf{M}_i$, the reconstructor 
$f_\theta$ is trained to predict the clean images $\mathbf{x}_i$ from the aliased ones. In this case, we use a U-Net~\cite{ronneberger2015u}  as the reconstructor trained between magnitude aliased and clean images.
The learned network U-Net is used as a reconstructor inside the subsequent mask selection algorithm. The framework alternates between learning the reconstructor and updating the masks.

\subsection{Greedy and Iterative Coordinate Descent Mask Selection Algorithms} \label{sec:greedy_icd_algorithm}
The greedy algorithm~\cite{gozcu2018learning} generates high-quality sampling patterns for each training scan. It is a framework proposed for specifying samples in k-space that minimize the reconstruction error given a particular reconstructor. 
In general, the greedy sampling algorithm can be used with various reconstructors. Our framework runs the greedy algorithm for each scan separately and thus, in the process produces scan-adaptive masks that are used during testing. We use a trained U-Net model as the reconstructor inside the mask selection algorithm and it uses $\ell_2$ norm as the error metric (as in \eqref{eq:opt}) for learning samples.

Starting with no sampled lines or only fixed low-frequency lines, at each step of greedy mask optimization, the k-space phase encode or line that gives the lowest reconstruction error is added to a particular mask. The algorithm keeps adding lines until the sampling budget is reached. The proposed iterative coordinate descent (ICD) scheme then further optimizes the mask iteratively by picking one line at a time in the current mask and moving it to the best new location in terms of the reconstruction error. Algorithm~\ref{alg:icd} summarizes the details of the main ICD algorithm. The ICD step is a novelty over prior work~\cite{gozcu2018learning}. The greedy and then ICD masks are obtained for each scan individually (patient and scan adaptive).

\begin{algorithm}[!t]
\caption{ICD Mask Optimization}
\label{alg::mb}
\begin{algorithmic}[1]
\Require Fully sampled k-space $\textbf{y}$ and corresponding forward operator $\mathbf{A}$, ground truth image $\mathbf{x}_{gt}$, reconstructor $f$,  loss function $L$, budget $B$, number of ICD iterations $N_{iter}$,  set of all possible line locations $\mathbf{S}$, set of locations of initial sampled lines $\mathbf{\Omega}_{\mathrm{initial}}$, initial mask $\mathbf{M}_{\mathbf{\Omega}_{\mathrm{initial}}}$
\State $\mathbf{\Omega} \leftarrow \mathbf{\Omega}_{initial}$
\For{$j=1:N_{iter}$}
\State $\{l_i\}_{i=1}^B\leftarrow$ entries in current $\mathbf{\Omega}$
\For{$i=1:B$}
    \State $\mathbf{\Omega}' = \mathbf{\Omega}  \setminus  l_i$ 
    \State $\mathbf{\Omega} \leftarrow \mathbf{\Omega}' \cup S^*$ where
    \begin{equation*}
        S^*=\underset{S \in \mathbf{S},\, S \notin \mathbf{\Omega}'}{\arg\min} 
        \, L(\mathbf{x}_{gt},f(\mathbf{A}^H \mathbf{M}_{\mathbf{\Omega}' \cup S} \mathbf{y}))
    \end{equation*}
    where $\mathbf{M}_{\mathbf{\Omega}' \cup S}$ is the operator sampling along lines at $\mathbf{\Omega}' \cup S$.
\EndFor
\EndFor
\State \textbf{return} $\mathbf{\Omega}$
\end{algorithmic} \label{alg:icd}
\end{algorithm}

\subsection{Training Reconstructors - U-Net and MoDL}
We used a U-Net~\cite{ronneberger2015u} as the reconstruction network inside the mask selection algorithms which are greedy~\cite{gozcu2018learning} and ICD algorithms. It is a relatively fast reconstructor inside the sampling optimization algorithms compared to say compressive sensing or iterative reconstruction.

After the final ICD masks are obtained, we also train
a state-of-the-art MoDL reconstruction network~\cite{aggarwal2018modl}. MoDL works with the following optimization problem:
\begin{equation} \label{eqmodl}
    \mathbf{x}_{rec} = \underset{\mathbf{x}}{\arg\min} \| \mathbf{A}\mathbf{x}-\mathbf{y}\|_2^2 + \lambda \|\mathbf{x}-D(\mathbf{x})\|_2^2,
\end{equation}
where $D(\mathbf{x})$ is a denoising neural network (e.g., a U-Net~\cite{ronneberger2015u} or DIDN network~\cite{yu2019deep}).
MoDL tackles~\eqref{eqmodl}
by introducing an auxiliary variable $\mathbf{z}$ and adopting an alternating algorithm as follows: 
\begin{align}
    \mathbf{x}_{n+1} &= \underset{\mathbf{x}}{\arg\min} \| \mathbf{A} \mathbf{x}-\mathbf{y}\|_2^2 + \lambda \|\mathbf{x}-\mathbf{z}_n\|_2^2\\
    \mathbf{z}_n &= D(\mathbf{x}_n)
\end{align}
The first subproblem can be solved using conjugate gradients (CG) based on the following normal equation:
\begin{equation}
    \mathbf{x}_{n+1} = (\mathbf{A}^H \mathbf{A} + \lambda \mathbf{I})^{-1} (\mathbf{A}^H \mathbf{y} + \lambda \mathbf{z}_n)
\end{equation}

The alternating scheme is unrolled for a few iterations and the denoising network is trained end to end in MoDL.

\subsection{Algorithm at Testing Time - Nearest Neighbor Search} \label{sec:nn_search}
Given our collection of scan-adaptive sampling masks obtained from the training process mentioned in the previous subsection, the task at the testing time is to estimate the locations of high-frequency samples in k-space (for Cartesian sampling) based on initially acquired low-frequency information. 
These high frequencies at test time are obtained directly from the pre-learned 
mask of the nearest neighbor in the training set. We perform the nearest neighbor search 
by comparing the 
reconstruction from the low-frequency k-space of a test scan with the low-frequency reconstruction of each training case.
We found the nearest neighbor by comparing the Euclidean distance between such images as follows:
\begin{equation}
   d_i =  \|\mathbf{A}^H_{\mathrm{test}} \mathbf{y}^{\mathrm{lf}}_{\mathrm{test}} - \mathbf{A}^H_{\mathrm{train_i}} \mathbf{y}^{\mathrm{lf}}_{\mathrm{train_i}} \|_2,
   \label{eq:nn}
\end{equation}
where $\mathbf{y}^{\mathrm{lf}}_{\mathrm{test}}$ is the low-frequency part of testing k-space with zeros at high frequencies, $\mathbf{y}^{\mathrm{lf}}_{\mathrm{train_i}}$ is the corresponding version for the $i$th training k-space, and $\mathbf{A}^H_{\mathrm{test}}$ and $\mathbf{A}^H_{\mathrm{train_i}}$ are the adjoints of the fully-sampled MRI forward operators for the test and $i$th training scans, respectively.

\section{Experiments and Results}
\label{sec3}
\subsection{Implementation Details}
We used the fastMRI multi-coil knee dataset~\cite{zbontar2018fastmri} for our experiments. The sensitivity maps were estimated from low-frequency k-space using the ESPIRIT Calibration approach~\cite{pruessmann1999sense}. The image size was $640 \times 368$ with $15$ coils for each k-space and the studies were performed at a sampling acceleration factor of 4x. The images were cropped to the central $320 \times 320$ region as it contains most of the useful contents. The reconstruction networks (U-Net) were trained between real-valued or magnitude (single channel) aliased and clean pairs of images for 100 epochs. Adam optimizer~\cite{kingma2014adam} was used for training the network with a learning rate of $10^{-3}$ and batch size of 1. The number of slices used for training and validation were 1200 and 300, respectively. 
We used normalized mean squared error (NMSE) as the metric for training our network which can be given by
\begin{equation}
    \text{NMSE} = \frac{\|\mathbf{x}_{\text{gt}}-\mathbf{x}_{\text{recon}}\|^2_2}{\|\mathbf{x}_{\text{gt}}\|^2_2},
\end{equation}
where $x_{\text{gt}}$ and $x_{\text{recon}}$ are the ground truth and reconstructed images, respectively.

We followed the training pipeline in Figure~\ref{fig:maintraining} with one pass of the ICD algorithm. The greedy mask optimization started with a low-frequency mask with $30$ lines initialized in the center and the rest were added using the search algorithm.
Once all ICD masks are generated, we post-train a two-channel U-Net and a MoDL reconstructor using these masks. The two channels of the network are the real and imaginary parts of the ground truth and aliased images. We also trained a U-Net and MoDL network for reconstructing from variable density random sampling (VDRS) masks.

\begin{figure}[h]
    \centering
    \includegraphics[width=0.9\linewidth]{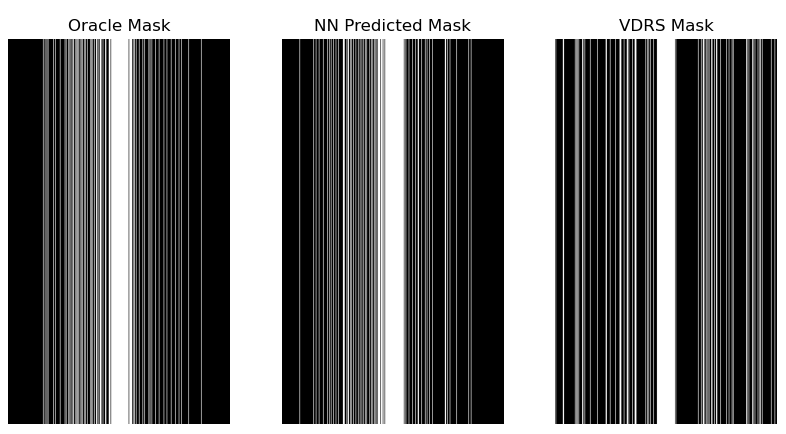}
    \caption{Some masks used for reconstruction.}
    \label{fig:allmasks}
\end{figure}

During testing time, we compared the reconstruction accuracy of our obtained ICD masks with the baseline VDRS mask shown in Figure~\ref{fig:allmasks}. The masks generated by the ICD algorithm during training were then selected by nearest neighbor search at test time. 

\subsection{Comparing Scan-adaptive with Population-adaptive Mask}
\begin{figure}[h]
\centering
\includegraphics[width=0.9\linewidth]{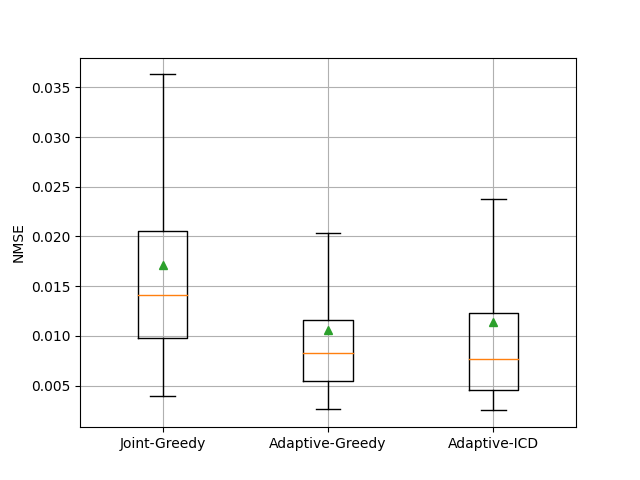}
\caption{Comparing patient and population-adaptive masks using U-Net reconstructor for 859 scans. The means and medians for different masks are indicated by green dots and orange lines.}
\label{fig:greedy_adaptive_vs_joint}
\vspace{-0.2in}
\end{figure}

Here, we compare the performance of the sampling patterns obtained by our scan-adaptive mask optimization approach to a learned population-adaptive sampling pattern. We obtained a population-adaptive greedy mask by optimizing the mask over the full training dataset and compared it with scan-adaptive greedy and the ICD masks. For testing, we trained a separate U-Net (single channel), to map aliased images generated with all three optimized masks, to ground truth images.
Figure~\ref{fig:greedy_adaptive_vs_joint} shows the NMSE of the reconstructed images obtained using these patient and population-adaptive masks on 859 
scans. 
We find that the scan-adaptive ICD and greedy masks gave better performance than the population-adaptive mask in terms of mean and median NMSE. The median ICD mask NMSE was somewhat better than with greedy masks.

\begin{table*}[!htb]
\centering
\begin{tabular}{|c|c|c|c|c|c|c|}
\hline
& VDRS with U-Net & \textbf{NN with U-Net} & Oracle with U-Net  & VDRS with MoDL & \textbf{NN with MoDL}  & Oracle with MoDL\\
 \hline
\textbf{NMSE}  & 0.076 & \textbf{0.049} & 0.047  & 0.036 & \textbf{0.034}  & 0.032 \\ \hline
\textbf{SSIM}  & 0.779 & \textbf{0.827} & 0.832  & 0.881 & \textbf{0.901}  & 0.906 \\ \hline
\textbf{HFEN}  & 0.896 & \textbf{0.823} & 0.816  & 0.782 & \textbf{0.728}  & 0.709 \\ \hline
\end{tabular}
\caption{Mean NMSE, SSIM, and HFEN values for reconstructed images using two-channel MoDL and U-Net reconstructors for oracle ICD masks, nearest neighbor (NN) ICD masks (our approach), and VDRS masks for 144 test cases. The U-Net was used and jointly adapted during training of the sampling masks whereas the MoDL network was post-trained on the generated ICD masks.}\label{tab:nmse}
\end{table*}

\begin{figure*}
    \centering
    \begin{subfigure}{0.9\linewidth}
        \centering
        \includegraphics[width=0.7\linewidth]{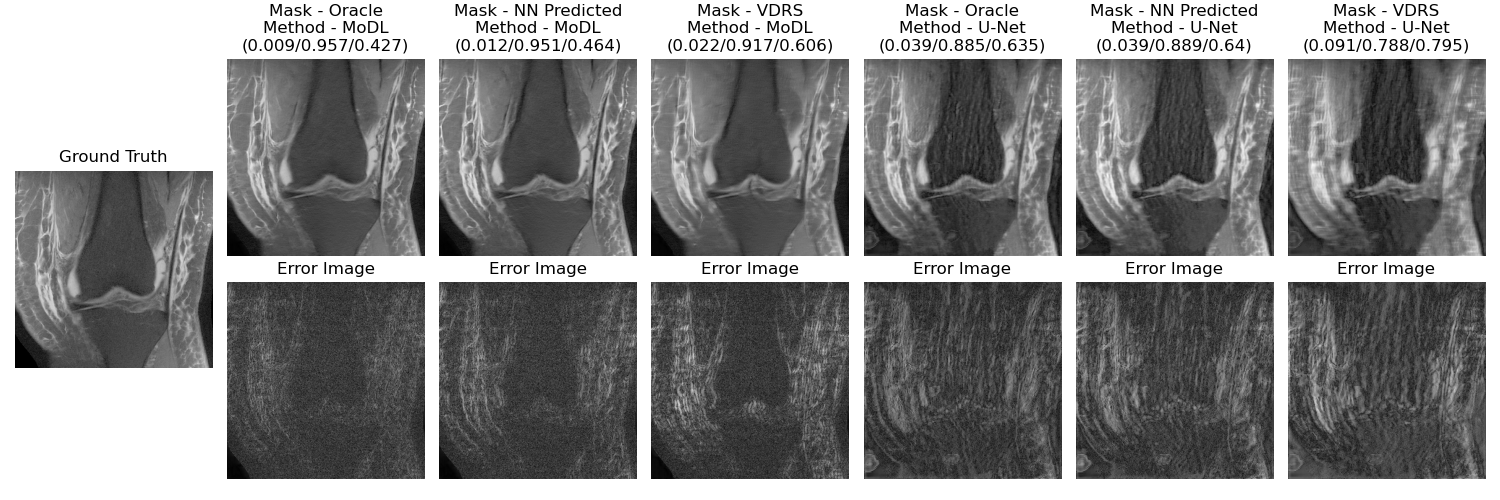}
        \label{fig:icd_recon1}
    \end{subfigure}
    
    \begin{subfigure}{0.9\linewidth}
        \centering
        \includegraphics[width=0.7\linewidth]{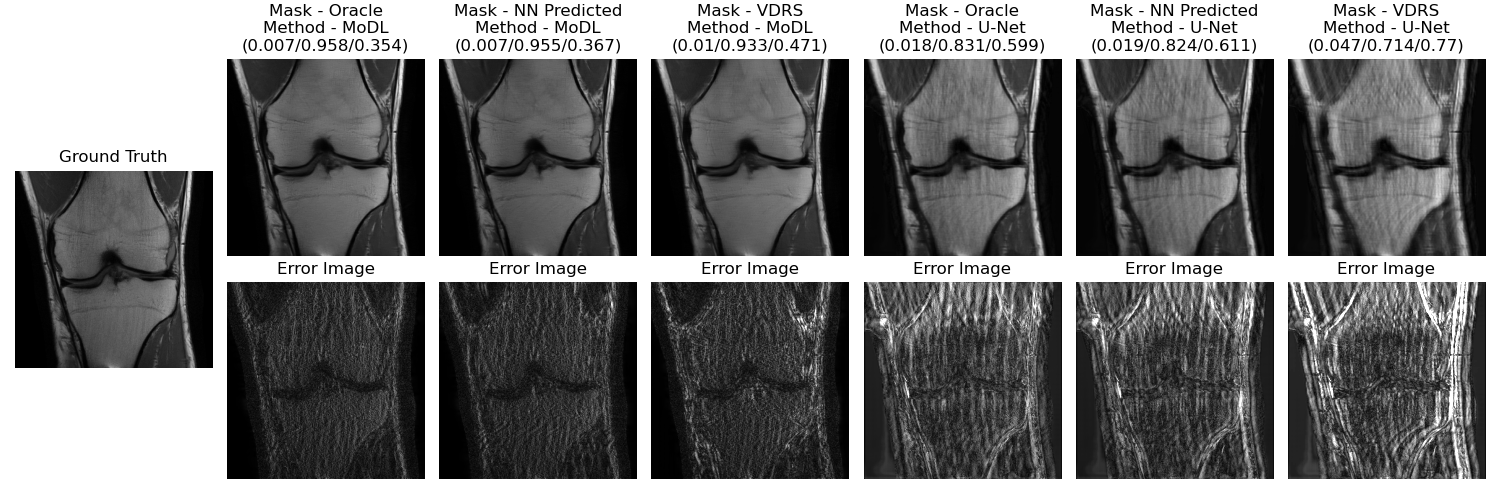}
        \label{fig:icd_recon2}
    \end{subfigure}
    \caption{Reconstructed and (magnitude) error images for different combinations of masks and reconstructors. Reconstruction metrics shown are (NMSE/SSIM/HFEN)}
    \label{fig:icd_recon}
    \vspace{-0.2in}
\end{figure*}

\subsection{Mask Prediction using Nearest Neighbor}

We use the scan-adaptive masks (obtained on the training set) generated by the ICD algorithm for the nearest neighbor search at test time. The reconstruction from the low-frequency part of k-space of a test case was compared with the reconstruction from the low-frequency part of all the training k-spaces and the Euclidean distance between them were noted (as in \eqref{eq:nn}). The nearest neighbors were found based on smallest Euclidean distance.

~\ref{tab:nmse} shows the average NMSE, SSIM, and HFEN values over the test cases for different combinations of reconstructors and masks. Figure~\ref{fig:icd_recon} shows reconstructed images for these combinations of different reconstructors and masks along with the error images for two different scans in the testing set.

Our results indicate better test-time reconstruction quality using an oracle (optimized directly using ICD for the test scan assuming known full k-space) and nearest neighbor sampling mask as compared to the conventional variable density random sampling (VDRS) mask in terms of NMSE, SSIM, and HFEN metrics, and visual quality (Figure~\ref{fig:icd_recon}). The test-time reconstructors used were the trained U-Net and MoDL networks.


Since our training optimizes both the sampling pattern and the reconstructor alternatingly, it provided us with optimized sampling patterns at 4x acquisition acceleration 
and the corresponding reconstructor as well. We used a U-Net reconstructor during training which is in general much faster than conventional iterative algorithms that take more time to run and iterations to converge. Our scheme is scan-adaptive, so it yields a pattern that is suited to each patient scan. In this work, we found the closest-looking neighbors based on Euclidean distance at test time to get the optimized ICD masks. For future work, we plan to learn a metric of image similarity that corresponds with eventual reconstruction quality (with sampling pattern obtained on the similar neighbor).

\section{Conclusion}
We proposed a scan-adaptive 
MRI subsampling method that estimates the high-frequency sample locations from low-frequency k-space information using sampling patterns optimized on neighbors in a training set. The proposed approach was tested on the fastMRI knee multi-coil dataset at 4x acceleration. 
In the initial training phase, our algorithm adaptively finds sampling patterns (using greedy and ICD algorithms) for each scan/slice together with a learned deep reconstructor. The reconstructor and sampling masks were optimized using the MSE metric. The learned scan-adaptive sampling patterns are used at test time via neighbor search, and were shown to have better accuracy than population-adaptive learned sampling patterns. 
Future work would involve using a learned CNN to predict best neighbors or the optimal sampling pattern and testing the approach at higher undersampling factors.

\section{Acknowledgments}
The authors acknowledge Drs. Jeffrey Fessler and Nicole Seiberlich from the University of Michigan, Shijun Liang from Michigan State University, and Zhishen Huang from Amazon Inc., for 
discussions.

\bibliographystyle{IEEEbib}
\bibliography{references}

\end{document}